# Formal Verification Of A Shopping Basket Application Model Using PRISM

Patrick Mukala
Departement of Computer Science, University of Pisa

**ABSTRACT**

Formal verification is at the heart of model validation and correctness. With model checking, invaluable realizations have been accomplished in software engineering and particularly in software development. By means of this approach, complex applications can be simulated and their performance forecasted in light with requirements at hands and expected performance. In this short paper we present the results of a simulation using Prism Model Checker for a Shopping Basket Application Model. Applied on a modified model from a projected process model, the objective is to simulate the behavior of shoppers as they go through a number of defined states of the shopping process and express accessibility and reachability through a number of defined properties.

*Corresponding Author:*

Patrick Mukala,
Departement of Computer Science,
University of Pisa,
Largo Pontecorvo 5, 56127 Pisa PI
Email:patrick.mukala@gmail.com, mukala@di.unipi.it

## 1. INTRODUCTION

It can be evidently proved that the value of model checking in automatically verifying correctness properties of finite-state systems is invaluable. A number of advantages for using model checking over other approaches based on simulation, testing, and deductive reasoning are extensively outlined in the literature such as the work in [6] and [7]. From complex hardware and software applications designs, model checking is used to verify required properties as specified and some of these include safety requirements such as the absence of deadlocks or their presence thereof as well as a range of similar critical states that can potentially cause the system to crash. Other requirements include satisfying correctness, liveness and persistence and product properties [7]. The most notable attractive particularity of model checking is the possibility to automatically perform verifications and offer counterexamples in case a model fails to satisfy a property serving as indispensable debugging information. This is performed through a number of model checking techniques and tools [8][9]. Some of these tools include BLAST, ROMEO, NusMV, SPIN, PAT, TAPS, PRISM and UPPAAL.

    The objective in this short document is to report on the use of one of the tools, namely Prism, through a case study as required for this course. Our case study translates a scenario used in [1] for process mining into a Discrete-Time Markov Chain (DTMC) Model that can be verified in Prism. Prism supports a number of models including discrete-time Markov chains (DTMCs), continuous-time Markov chains (CTMCs), Markov decision processes (MDPs), probabilistic automata (PAs) and probabilistic timed automata (PTAs). More details about these models can be found in [2].

    The scenario from which stems our case study in [1] deals with using process mining [9] to analyze systems audit trail. An audit trail is a record of all events that take place in a system and across a network, i.e., it provides a trace of user/system actions so that security events can be related to the actions of a specific individual or system component [1]. The scenario can be described as follows: A website specializing in selling products is considered. It is assumed that all registered users are assigned a shopping basket that can be edited at any time. While shopping, the user selects products and puts them in the basket and these remain in the basket unless they are removed by the user even after logging out. This implies that the user basket's status is saved and is retrieved when the user enters the website again. We choose a number of inherent user activities as depicted in Figure 1 below.

From this figure, a log of the audit trail can be retained to understand the usage of the web site and user movements. Considering that such a log be WOK = {"Enter, Select Product, Add to Basket, Cancel Order", "Enter, Select Product, Remove from Basket, Cancel ", "Enter, Select Product, Add to Basket, Continue Shopping, Select Product, Remove from Basket, Continue Shopping, Select Product, Add to Basket, Proceed to Checkout, Fill in Delivery Info, Fill in Payment Info, Provide Password, Process Order, Finish Checkout", "Enter, Select Product, Remove from Basket, Proceed to Checkout, Fill in Payment Info, Fill in Delivery Info, Provide Password, Process Order, Finish Checkout"}.

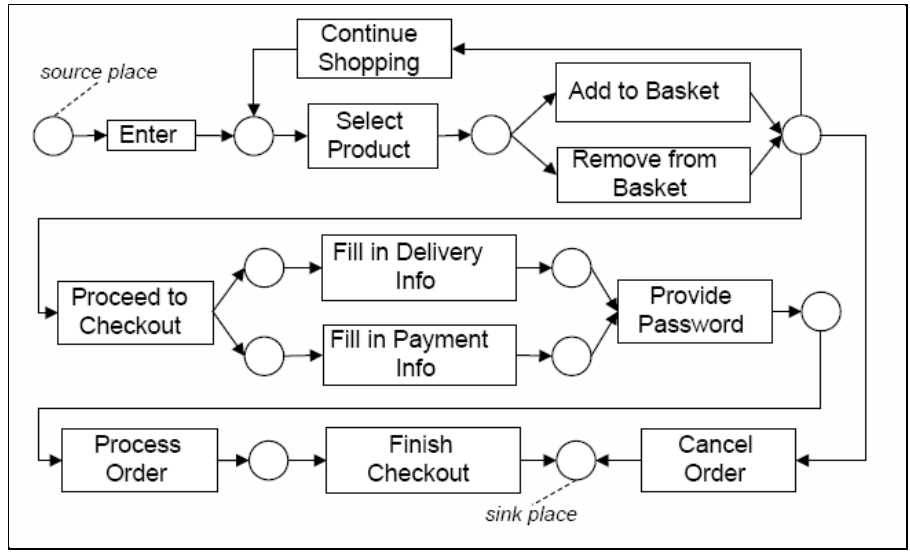

**Figure 1 : Process execution Flow for Online Shopping Basket Activities**

This information gives an idea of how users perform their actions online while using the services but does not predict the performance of the system if implemented. For the sake of our assignment, we derive a DTMC Model from this scenario as given in the following section and verify it in Prism.

The rest of the paper is structured as follows: in section 2, we succinctly talk about DTMCs by formally defining them, in section 3 we introduce and describe our model, while in section 4 we set a number of properties for verification and give the results as interpreted in section 5 and conclude in section 6.

## 2. DISCRETE TIME MARKOV CHAINS MODELS

In order to explain these models, we consider two definitions as follows.

**Definition 1** [2]: Formally, a DTMC D is a tuple $(S, s_{init}, P, L)$ where:
- S is a finite set of states ("state space")
- $s_{init} \in S$ is the initial state
- $P : S \times S \to [0,1]$ is the transition probability matrix where $\Sigma_{s' \in S} P(s,s') = 1$ for all $s \in S$
- $L : S \to 2^{AP}$ is function labeling states with atomic propositions

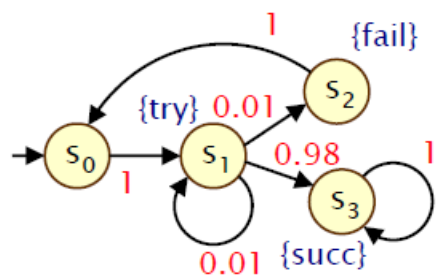

**Definition 2:** The stochastic process $X = (X_0, X_1,...)$ that takes values in some countable set $S$ is a discrete **Markov chain** if it satisfies the Markov property:

$$P(X_n = s | X_0, X_1,...X_{n-1}) = P(X_n = s | X_{n-1}),$$

for all $n > 0$, and $s$ in $S$.

This Markov property says essentially that the probability that the chain will visit state $s$ at step $n$ given all the past history is equal to the probability of visiting state $s$ at step $n$ given only the current state $X_{n-1}$ (at time $n-1$). The conditional probability $P(X_n = s | X_{n-1} = w)$ is referred to the one step transition probability of the Markov chain from state $s$ to state $w$ at step $n$. If in addition, the transition from one state to the other does not depend on the step $n$, that is

$$P(X_n = s | X_0, X_1,...X_{n-1}) = P(X_n = s | X_{n-1}),$$

for all $n$, then we say that the Markov chain is **stationary** or **homogeneous**.

For a stationary Markov chain, it is sufficient to specify the *one-step transition probability*, $p_{ij} = P(X_n = j | X_{n-1} = i)$. The square matrix **P** whose elements are the $p_{ij}$'s is called the *one-step transition matrix*, or just the *transition matrix* of the Markov chain.

The *n-step transition probabilities*, $p_{ij}^{(n)}$, are defined by $p_{ij}^{(n)} = P(X_n = j | X_0 = i)$. That is, $p_{ij}^{(n)}$ is the probability that the chain visits state $j$ at step $n$, given that it is initially at state $i$ at step $0$. Note that $p_{ij}^{(1)}$ is simply $p_{ij}$.

We may derive the following relationship as follows.

$$p_{ij}^{(n)} = P(X_n = j | X_0 = i)$$
$$= \sum_k P(X_n = j, X_{n-1} = k | X_0 = i) = \sum_k P(X_n = j | X_{n-1} = k, X_0 = i) P(X_{n-1} = k | X_0 = i)$$
$$= \sum_k P(X_n = j | X_{n-1} = k) P(X_{n-1} = k | X_0 = i) = \sum_k P(X_1 = j | X_0 = k) P(X_{n-1} = k | X_0 = i)$$
$$= \sum_k p_{kj} p_{ik}^{(n-1)}$$

In conclusion, $p_{ij}^{(n)} = \sum_k p_{ik}^{(n-1)} p_{kj}$, for all states $i$, and $j$ and for all steps $n$. In matrix notation, this is equivalent to

$$\mathbf{P}^{(n)} = \mathbf{P}^{(n-1)}\mathbf{P},$$

where $\mathbf{P}^{(n)}$ is the matrix whose elements are the *n-step transition probabilities*

With DTMC, a transition corresponds to the advance of a single time-unit. Therefore, the underlying time domain is thus discrete because the present moment refers to the current state and the next moment corresponds to the immediate successor state. Simply put, with these models the system behavior can be observed at the time points 0,1,2,.... of real-time constraints in asynchronous systems by means of a discrete-time domain. A discrete time domain conceptually allow for transition systems to be modeled as timed systems where each action is assumed to last for a single time unit. More general delays can be modeled by using a dedicated unobservable action, τ (for tick), say. The fact that action α lasts k > 1 time units may be modeled by k−1 tick actions followed (or preceded) by α. This approach typically leads to very large transition systems [7].

## 3. THE MODEL : A DTMC MODEL FOR SHOPPING BASKET APPLICATION

A DTMC model boasts as depicted in Figure 2 boasts 14 states that are annotated to signify the corresponding state number and its description. In this model, the user starts with being in the *BrowseShop* state with a 30% probability to continue just browsing through announcements and product detailing before deciding whether to continue just browsing or logging in and making purchases, with the latter's occurrence estimated at 70 % probability. The next state is *LoggedIn* signaling the user's presence and where he/she is 100% sure to have permission to select products in the *SelectProduct* state where he/she can either remove products from the shopping basket or add more at split chance. If the user decides to delete from the already selected basket at the *DelFrBasket* state, the next steps would require them to either continue shopping for other alternative products through *KeepShopping* with only 35% probability where he/she will be redirected to select new products through *SelectProduct* or he/she can decides to cancel the order altogether at the

*CancelOrder* state with 65 % probability at which point they are required to log out through the *LoggedOut* state.

On the other hand, if the user decides after selecting product to add to the basket through the *AddToBasket* state, he/she has a fifty-fifty choice to either start checking and move to the *StartCheckout* state or continue shopping through *KeepShopping* at which point he/she is redirected and taken back to *SelectProduct* to choose new products. At *StartCheckout* state, the user has a possibility at equal probabilities to either fill in only payment information for the selected products and hence move to *FillPaymentInfo* state or choose an alternative option for providing payment details with a possibility for delivery at *FillInDeliveryInfo*. If the user gets to the *FillPaymentInfo* state, he/she can either proceed to the next step for financial credentials authentication at *Authenticate* or go back to *StartCheckout* if there is a change of heart and the need to rather choose the option with possibility of delivery at *FillInDeliveryInfo*. Similarly, if the initial choice was to request and invoke the delivery service through the *FillInDeliveryInfo,* the user has a 50 % choice to either proceed to *Authenticate* or return to *StartCheckout.*

At *Authenticate,* three possibilities occur. The user is either redirected to the previous step (in case of incorrect credentials) with 25% probability for both *FillInDeliveryInfo* and *FillPaymentInfo* each and a 50% probability to move to process the order at *ProcessOrder* . From this state, he/she is moved to the nest step called *CompleteCkeckout* for confirmation details and printing thereof and then the user can conclude the shopping transactions through the last state called *LoggedOut.*

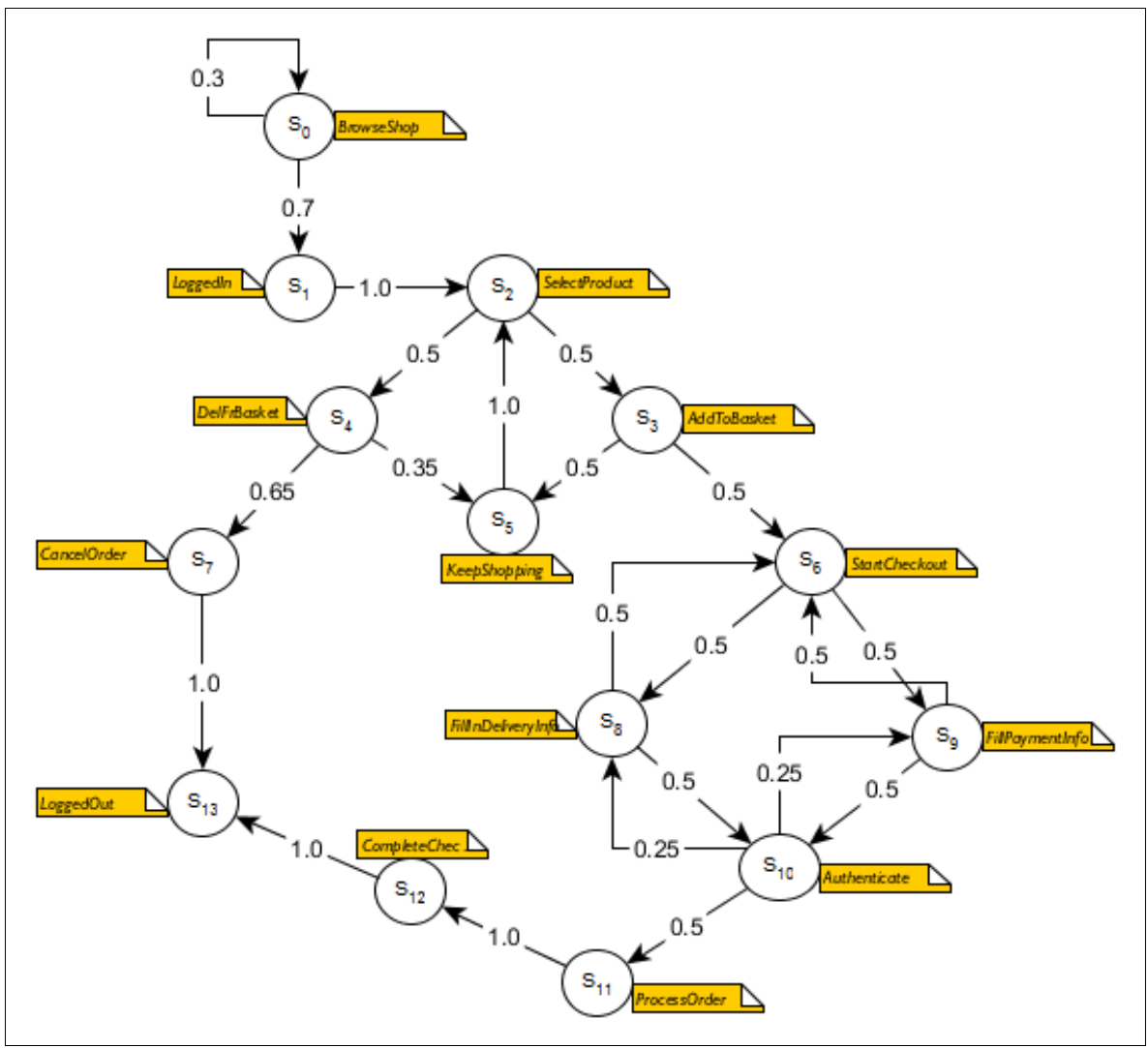

Figure 2 : A DTMC Model for Shopping Basket Agent

cancellation

// Model Checking a shopping basket model
dtmc

module shopper

    // declaring local states from the shopping basket model
    s : [0..13] init 0;

    [] s=0 -> 0.3:(s'=0) + 0.7:(s'=1);
    [] s=1 -> 1.0:(s'=2);
    [] s=2 -> 0.5:(s'=3) + 0.5:(s'=4);
    [] s=3 -> 0.5:(s'=5) + 0.5:(s'=6);
    [] s=4 -> 0.65:(s'=7) + 0.35:(s'=5);
    [] s=5 -> 1.0:(s'=2);
    [] s=6 -> 0.5:(s'=8) + 0.5:(s'=9);
    [] s=7 -> 1.0:(s'=13);
    [] s=8 -> 0.5:(s'=10)+ 0.5:(s'=6);
    [] s=9 -> 0.5:(s'=10)+ 0.5:(s'=6);
    [] s=10 -> 0.25:(s'=8) + 0.25:(s'=9) + 0.5:(s'=11);
    [] s=11 -> 1.0:(s'=12);
    [] s=12 -> 1.0:(s'=13);
    [] s=13 -> true;

endmodule

## 4. PROPERTIES FOR VERIFICATION

In order to verify this model in Prism, we set a number of properties as threshold for verification. The idea is that we want to get a picture of the application performance with regards to a number of expectancy factors. These include understanding and simulating at what extent an application user will be able to successfully complete shopping to keep shopping after deleting from Basket, to eventually come back and include delivery information after initially ignoring the choice, to cancel the order, to eventually buy, after deleting some products from the basket . Additionally, one would like to estimate that given this scenario, what would be the probability that a user would eventually reach the checkout state, give erroneous authentication financial credentials, will log out successfully, would select either simply paying or requesting delivery as well, and finally, the probability that a user would delete selected products.

Based on these requirements, we formulate ten properties as follows that can be transformed into model properties and verify them in Prism as required.

- → *$P_1$: Probability of eventually reaching state 12, which indicates that the user has completed successfully shopping and bought products is less than 0.7* :

  P< 0.7 [ F (s = 12)]

- → *$P_2$: Probability to keep shopping after deleting from Basket shall be less than 0.5*

  P< 0.5 [ F (s = 5) { (s=4)}]

- → *$P_3$: Probability to eventually come back and include delivery information after initially ignoring the choice is greater or equal to 0.5*

  P >= 0.5 [F (s = 9) {(s = 8)}]

- → *$P_4$: Probability to cancel the order shall be less than 0.8*

P< 0.8 [ F (s = 7)]

→ $P_5$: Probability to eventually buy, after deleting some products from the basket shall be more than 0.5

P > 0.5 [ F (s = 12) { s = 4}]

→ $P_6$: What is the probability that a user would eventually start checkout?

P=? [ F (s = 6)]

→ $P_7$: What is the probability that a user would give erroneous authentication for financial credentials?

P? [ F (s = 9) {(s =10) }]

→ $P_8$: What is the probability that a user will log out successfully?

P=? [ F( s = 13)]

→ $P_9$: What is the probability that a user would select either simply paying or requesting delivery as well?

P=? [ F((s = 8)|(s = 9))]

→ $P_{10}$: What is the probability that a user would delete selected products?

P=? [ F (s = 4)]

## 5. PROPERTIES FOR VERIFICATION

| Properties | Verification Results |
|---|---|
| → Property 1: P<0.7 [ F (s=12) ] | Number of states satisfying P<0.7 [ F (s=12) ]: 7<br>**Result: true (property satisfied in the initial state)** |
| → Property 2: P<0.5 [ F (s=12) {(s=4)} ] | Number of states satisfying P<0.5 [ F (s=12) {(s=4)} ]: 7<br>**Result: true (property satisfied in all filter states)** |
| → Property 3: P>=0.5 [ F (s=9) {(s=8)} ] | Number of states satisfying P>=0.5 [ F (s=9) {(s=8)} ]: 4<br>**Result: true (property satisfied in all filter states)** |
| → Property 4: P<0.8 [ F (s=7) ] | Number of states satisfying P<0.8 [ F (s=7) ]: 12<br>**Result: true (property satisfied in the initial state)** |
| → Property 5: P>0.5 [ F (s=12) {(s=4)} ] | Number of states satisfying P>0.5 [ F (s=12) {(s=4)} ]: 7<br>**Result: false (property not satisfied in all filter states)** |
| → Property 6: P=? [ F (s=6) ] | **Result: 0.4347821160949293** |
| → Property 7: P=? [ F (s=9) {(s=10)} ] | **Result: 0.3999999999966359** |
| → Property 8: P=? [ F (s=13) ] | **Result: 1.0 (value in the initial state)** |
| → Property 9: P=? [ F ((s=8)|(s=9)) ] | **Result: 0.4347821160949293** |
| → Property 10: P=? [ F (s=4) ] | **Result: 0.666666030883789** |

The visual representation of the results of the verification is depicted in the snapshot in Figure 3 below.

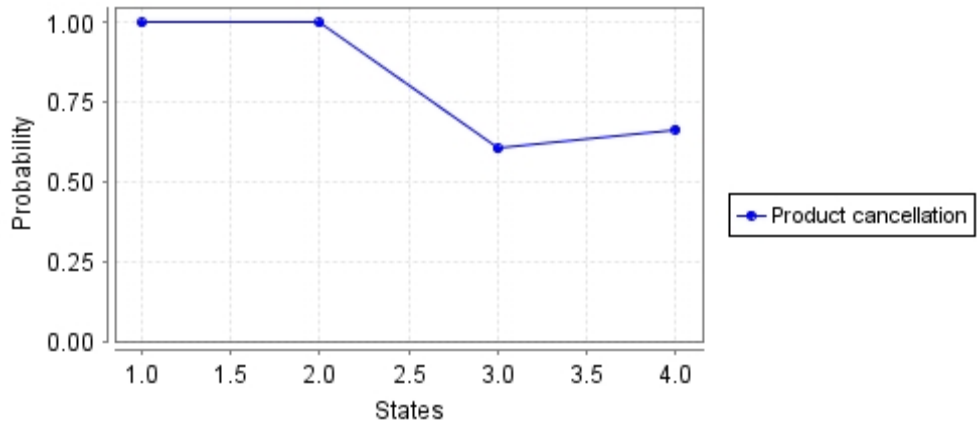

Figure 3 : Snapshot of Properties Validation

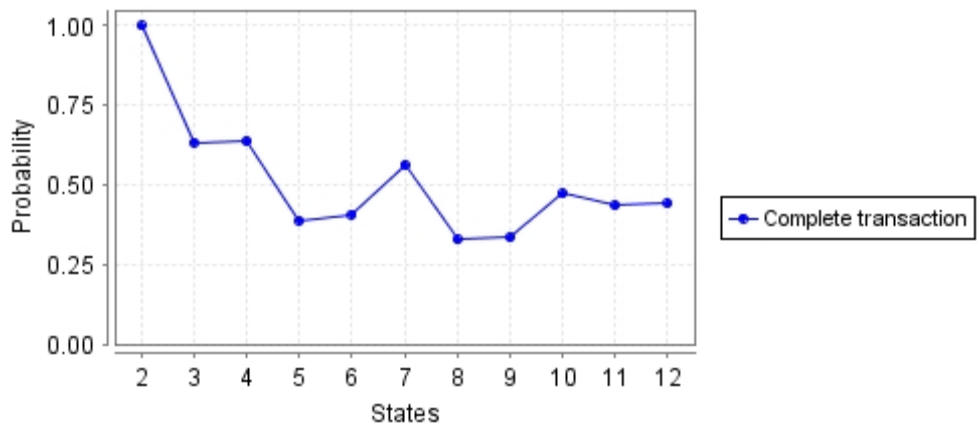

Figure 4 : Probability of Deleting A product from Basket

Figure 5 : Probability of Successfully Completing A transaction

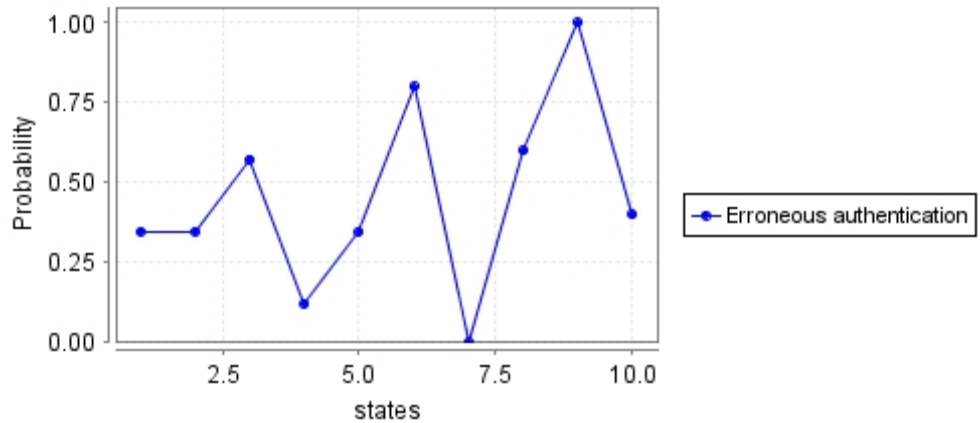

**Figure 6 : Probability of Supplying incorrect financial credentials**

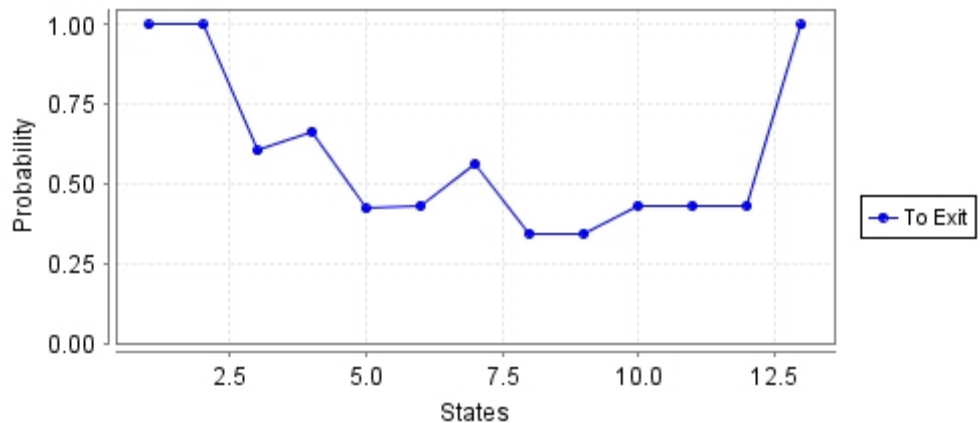

**Figure 7 : Probability of Interrupting or Completing a successful transaction and exitt**

As given by the snapshot in figure 3, the simulation this experiment suggests that 4 indicators can be successfully verified except from the expected probability to eventually buy, after deleting some products from the basket to be more than 50. The rest of properties are expected to give numerical values as depicted in Figures 4 to 7.

6. **CONCLUSION**

In this short paper, the objective was to simulate a DTMC model and verify a number of properties using Prism Model Checker. A case study was chosen to this end in order to facilitate this process. We constructed a DTMC model from a scenario on an online shopping basket environment and proposed a number of properties for verification as depicted in the previous section. We kept our case study is simple as possible and ensured the verification process demonstrated the fundamentals of model checking with PRISM.

**BIBLIOGRAPHY OF AUTHOR**

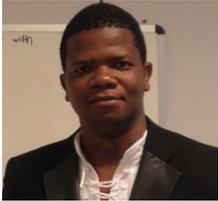
Patrick Mukala is currently a PhD Candidate in Computer Science at the University of Pisa in Italy. His doctoral work includes process and data mining in FLOSS repositories.